\documentclass[lettersize,journal]{IEEEtran}

\usepackage{amsmath,amsfonts} 
\usepackage{algorithmic}
\usepackage{algorithm}

\usepackage{array,ltablex,tabularx}
\usepackage{setspace} 
\usepackage{stfloats} 
\usepackage{booktabs} 

\usepackage{graphicx} 
\usepackage{subcaption} 
\usepackage{textcomp} 
\usepackage{xcolor} 

\usepackage{cite} 
\usepackage{url} 

\usepackage{newtxtext,newtxmath} 
\usepackage{mathtools} 
\usepackage{pifont} 

\usepackage{orcidlink} 

\usepackage{titlesec}
\titlespacing*{\section}{0pt}{*0.5}{*0.5} 
\titlespacing*{\subsection}{0pt}{*0.3}{*0.3}

\usepackage{verbatim} 

\usepackage{lipsum} 

\usepackage{balance} 

\begin{document}

\title{3D-HQAM Constellation Design and Performance Evaluation under AWGN}

\author{
Sukhsagar,~\IEEEmembership{Graduate Student Member, IEEE}%
    \thanks{Sukhsagar is with the Centre for Advanced Electronics and V. Bhatia is with the Department of Electrical Engineering, Centre for Advanced Electronics, Indian Institute of Technology Indore, Indore 453552, India (e-mail: phd2301191002@iiti.ac.in).},
    Nagendra Kumar,~\IEEEmembership{Senior Member,~IEEE}%
    \thanks{Nagendra Kumar is with the Department of Electronics and Communication Engineering, National Institute of Technology, Jamshedpur 831014, India (e-mail: kumar.nagendra86@gmail.com).}, 
    Ambuj Kumar Mishra,~\IEEEmembership{}%
    \thanks{Ambuj Kumar Mishra is with the Department of Mathematics, Institute of
Applied Sciences and 0Humanities, GLA University Mathura, Mathura 281406,
India (e-mail: ambuj.mishra@gla.ac.in).
    } Vimal Bhatia,~\IEEEmembership{Senior Member,~IEEE}%
    \thanks{Vimal Bhatia is with School of Electronic and Information Engineering,
Soochow University, Suzhou 215006, China, also with Škoda Auto University,
293 01 Mlada Boleslav, Czech Republic, and Faculty of
Informatics and Management, University of Hradec Kralove, 500 03 Hradec
Kralove, Czechia (e-mail: vbhatia@iiti.ac.in).
    } and
   Ondrej Krejcar,~\IEEEmembership{Senior Member,~IEEE}%
    \thanks{O. Krejcar is with Skoda Auto University, Na Karmeli 1457, 293 01 Mlada Boleslav (e-mail: ondrej.krejcar@uhk.cz).
    } }
    
\maketitle
\begin{abstract}
This paper proposes a simple and effective method for constructing higher-order three-dimensional (3D) signal constellations, aiming to enhance the reliability of digital communication systems. The approach systematically extends the conventional two-dimensional hexagonal quadrature amplitude modulation (2D-HQAM) constellation into a 3D-HQAM signal space, forming structured lattice configurations. To address the increased decision complexity resulting from a larger number of constellation points, a dimension reduction (DR) technique is introduced, allowing the derivation of closed-form symbol error probability (SEP) expressions under additive white Gaussian noise (AWGN)  conditions. Theoretical SEPs closely match simulation results, validating the accuracy of the proposed method. The minimum Euclidean distance (MED) of the 3D constellations shows a minimum increase of 12.14\% over 2D constellation for 8-HQAM, reaching up to 160.81\% for 1024-HQAM constellations. This significant improvement in MED leads to enhanced error performance. Therefore, the proposed 3D constellations are promising candidates for high-quality and reliable next-generation digital communication systems.
\end{abstract}

\begin{IEEEkeywords}
Digital communication, 3D signal constellations, HQAM, Dimensionality Reduction, MED, SEP, and AWGN.
\end{IEEEkeywords}

\section{Introduction}
\IEEEPARstart{T}{he} rapid growth of mobile services has intensified the need for advanced wireless communication systems capable of supporting higher data rates and improved transmission reliability. In such systems, digital modulation techniques play a critical role, where binary bit streams are mapped to real or complex-valued symbol sequences using signal constellations. Among the widely adopted two-dimensional (2D) constellations, quadrature amplitude modulation (QAM) and hexagonal QAM (HQAM) \cite{Sukhsagar2025accurate} are commonly employed due to their balance of spectral efficiency and implementation simplicity. A key limitation of conventional 2D modulation schemes is that, under a fixed transmit power constraint, increasing the modulation order reduces the minimum Euclidean distance (MED) between constellation points \cite{rugini2016symbol}. This reduction weakens signal robustness and requires a higher signal-to-noise ratio (SNR) for reliable demodulation. Additionally, higher-order 2D constellations introduce stricter radio frequency (RF) design constraints, increasing complexity and cost.

To overcome these limitations, advanced three-dimensional (3D) mapping techniques allow constellation points to be distributed in 3D space, thereby enhancing spatial degrees of freedom. This spatial expansion increases the MED under the same power constraint, resulting in improved error performance and higher system throughput for a given symbol error probability (SEP). MED is a crucial parameter in digital communication systems, as a greater MED enhances noise immunity and reduces the SEP. When the number of signal points is fixed and average power is normalized, the MED tends to increase with the dimensionality of the signal space. Therefore, 3D constellations offer stronger error resilience compared to 2D designs, making them well-suited for high-order modulation in the next-generation wireless communication systems. A lot of research on the construction of new multidimensional signal constellations has been reported. For instance, exact SEP for arbitrary 3D constellations has been derived in \cite{khabbazian2009exact}, while \cite{beko2012designing} proposed energy-efficient multidimensional constellation designs. Construction of lattice-based 3D constellations with closed-form SEP analysis was presented in \cite{kang2011construction}, and layered hexagonal designs improving MED were introduced in \cite{chen2015design}. Adaptive loading techniques for OFDM using 3D constellations are explored in \cite{chen2017three}, and index modulation in 3D space is advanced in \cite{huang2019quadrature}. Space-time 3D constellations enhancing diversity are proposed in \cite{zheng2019multi}, and 3D 64-QAM for multiple-input-multiple-output (MIMO)-visible light communication (VLC) systems is developed in \cite{guo2023superposed}. Partitioned 3D constellations for delta-sigma modulation were presented in \cite{zhao2024262144}. The integration of 3D constellations with chaos-based modulation in \cite{tan2024three} and polyhedral 3D 32-QAM for VLC systems in \cite{guo2025probabilistically} further demonstrate their broad applicability. These efforts highlight the growing relevance of 3D constellation design for improving performance in modern wireless systems.

Taking this into consideration, and to the best of our knowledge, no prior work in the literature has addressed the use of 3D constellations for the HQAM scheme. In this study, we propose a novel 3D constellation design for the HQAM scheme, referred to as 3D-HQAM. The primary contributions of this study are as follows:
\begin{itemize}
    \item To improve the communication dependability, a straightforward and methodical approach is suggested for extending 2D-HQAM into 3D space and creating structured 3D-HQAM lattice constellations.

    \item A dimension reduction (DR) technique is utilized to mitigate the heightened decision complexity of 3D-HQAM constellations, facilitating efficient projection onto a 2D plane while maintaining MED.
    \item Employing the nearest-neighbors (NNs) approach, a precise closed-form expression for the SEP under additive white Gaussian noise (AWGN) channels is developed and validated through simulations, showing close agreement with theoretical results.
\end{itemize}
\begin{figure}[htbp]
\centering


\begin{minipage}[b]{0.22\textwidth}
  \centering
  \includegraphics[width=3.2cm,height=2.4cm]{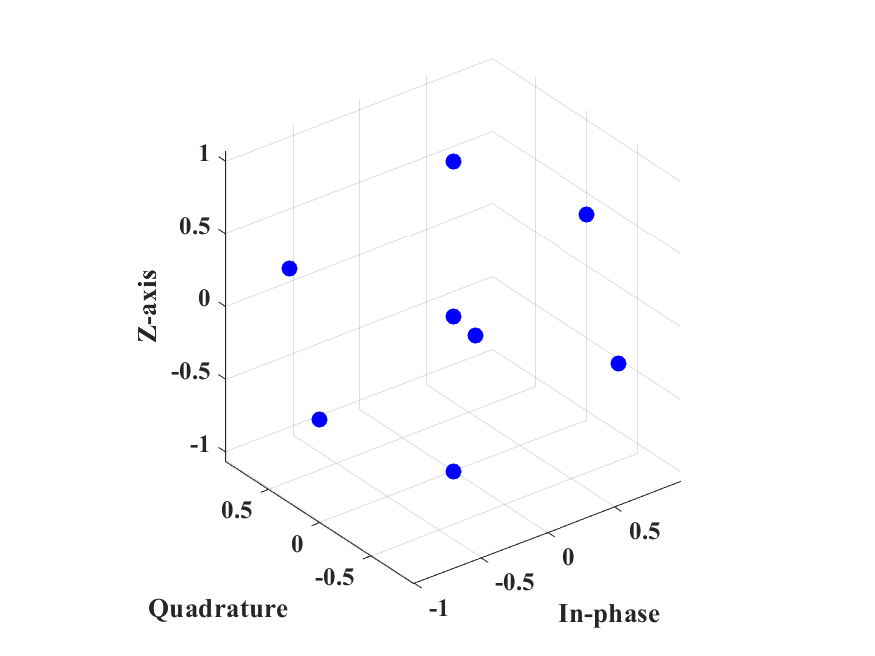}
  \caption*{M=8}
\end{minipage}
\begin{minipage}[b]{0.22\textwidth}
  \centering
  \includegraphics[width=3.2cm,height=2.4cm]{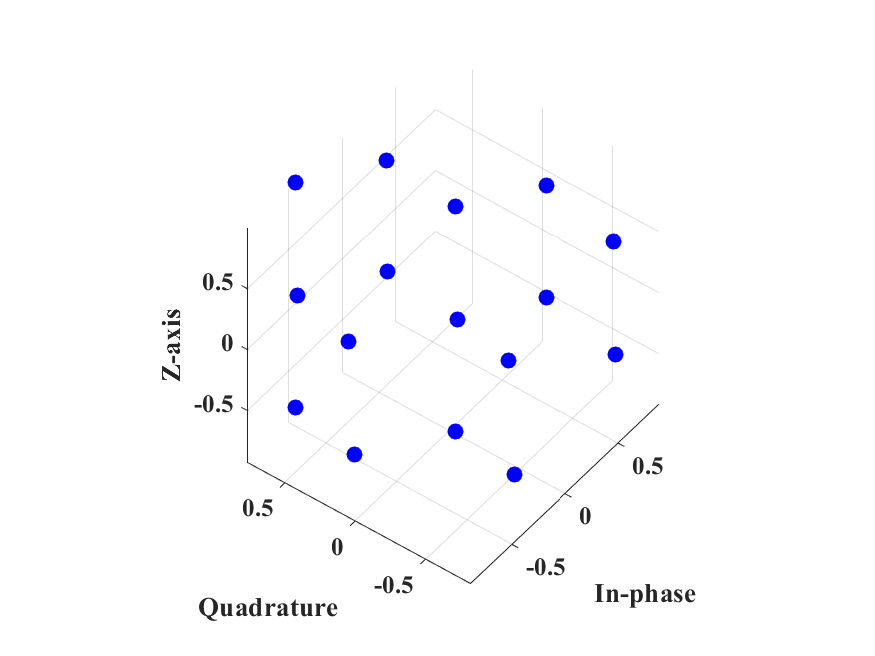}
  \caption*{M=16}
\end{minipage}
\begin{minipage}[b]{0.22\textwidth}
  \centering
  \includegraphics[width=3.2cm,height=2.4cm]{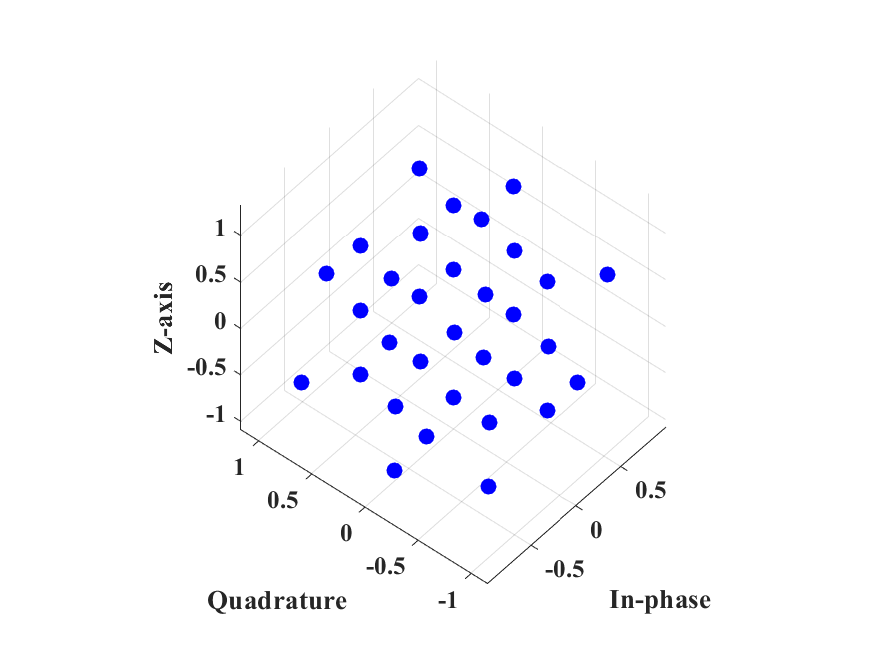}
  \caption*{M=32}
\end{minipage}
\begin{minipage}[b]{0.22\textwidth}
  \centering
  \includegraphics[width=3.2cm,height=2.4cm]{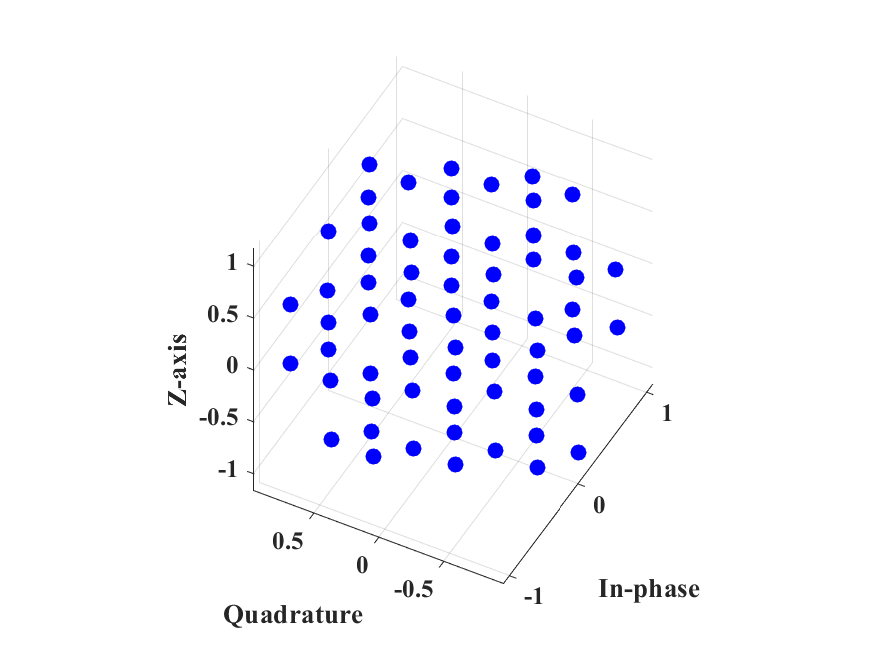}
  \caption*{M=64}
\end{minipage}
\vspace{1em} 

\begin{minipage}[b]{0.22\textwidth}
  \centering
  \includegraphics[width=3.2cm,height=2.4cm]{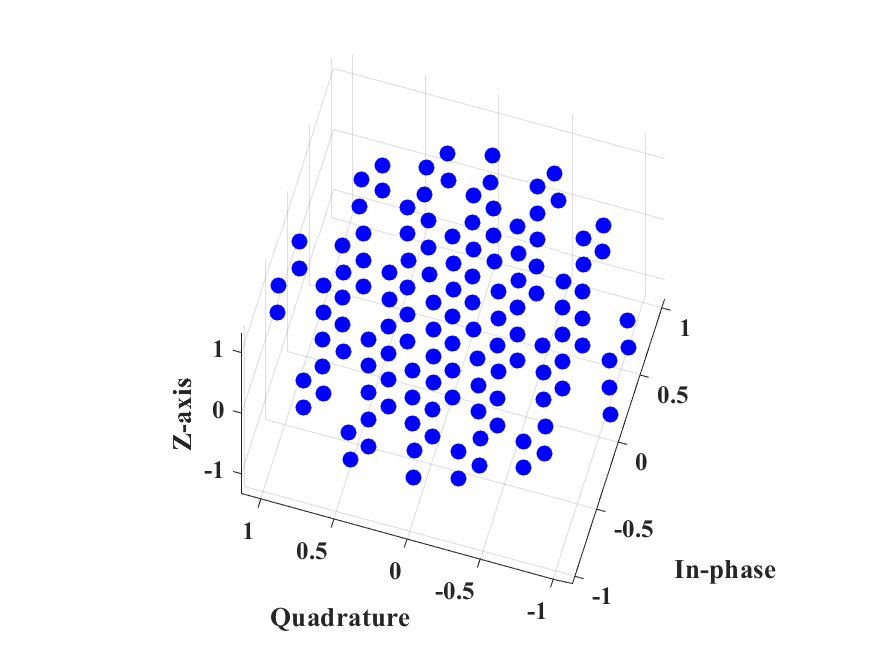}
  \caption*{M=128}
\end{minipage}
\begin{minipage}[b]{0.22\textwidth}
  \centering
  \includegraphics[width=3.2cm,height=2.4cm]{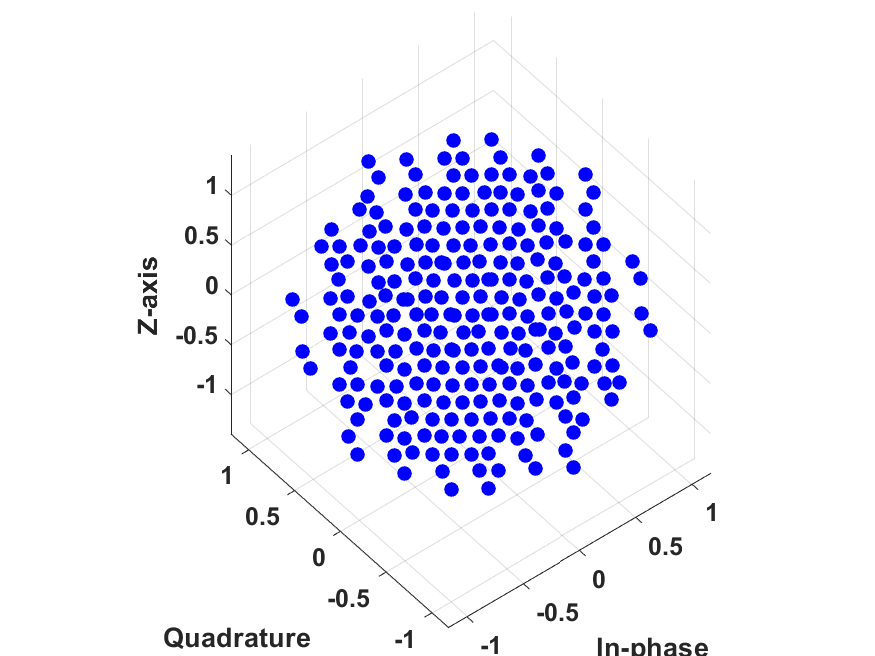}
  \caption*{M=256}
\end{minipage}
\begin{minipage}[b]{0.22\textwidth}
  \centering
  \includegraphics[width=3.2cm,height=2.4cm]{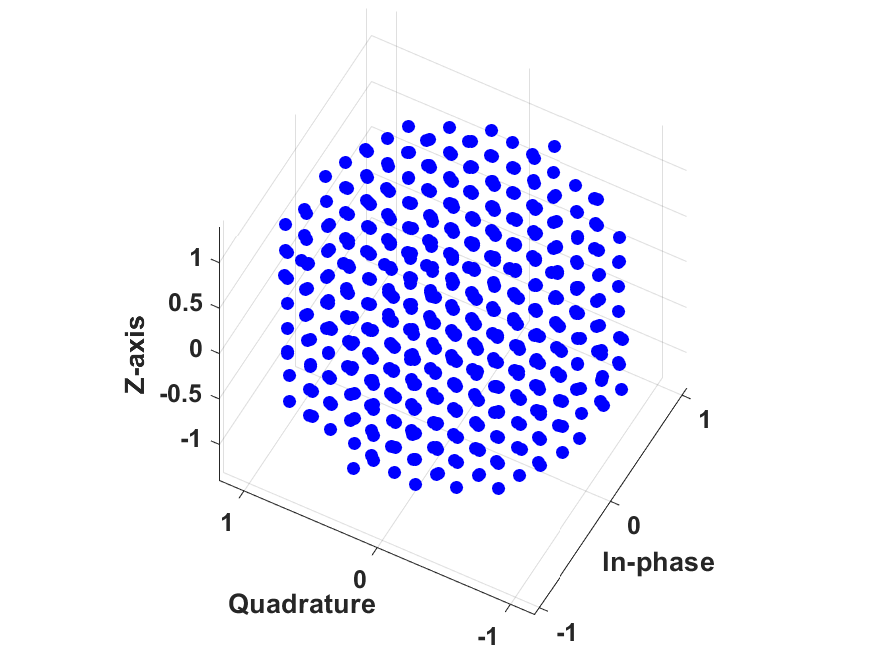}
  \caption*{M=512}
\end{minipage}
\begin{minipage}[b]{0.22\textwidth}
  \centering
  \includegraphics[width=3.2cm,height=2.4cm]{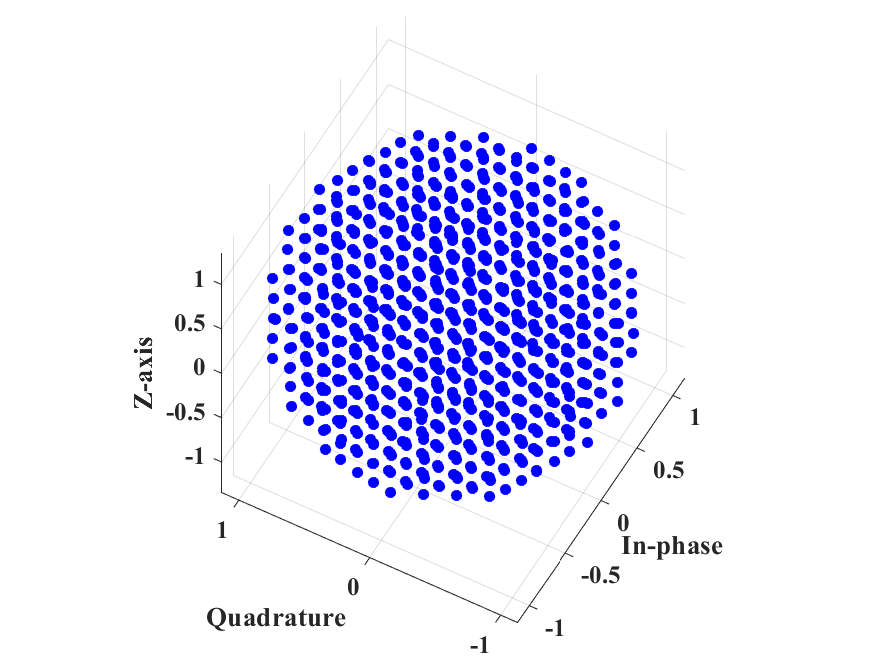}
  \caption*{M=1024}
\end{minipage}
\vspace{1em}
\caption{Scatter plot of 3D-HQAM constellations.}
\label{fig:1}
\end{figure}
\begin{figure}[htbp]
\centering


\begin{minipage}[b]{0.22\textwidth}
  \centering
\includegraphics[width=3.2cm,height=2.4cm]{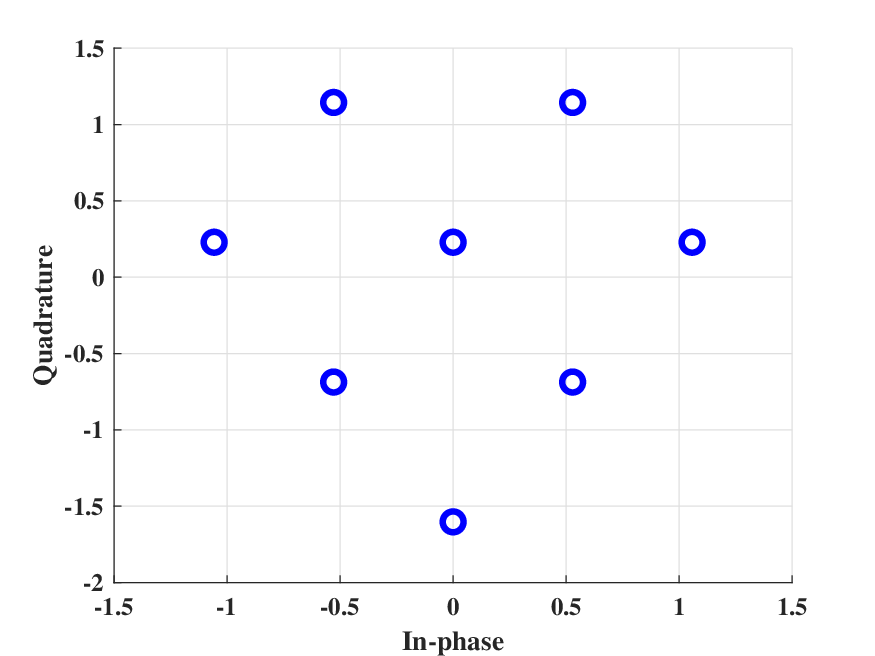}
  \caption*{Optimum, M=8}
\end{minipage}
\begin{minipage}[b]{0.22\textwidth}
  \centering
  \includegraphics[width=3.1cm,height=2.4cm]{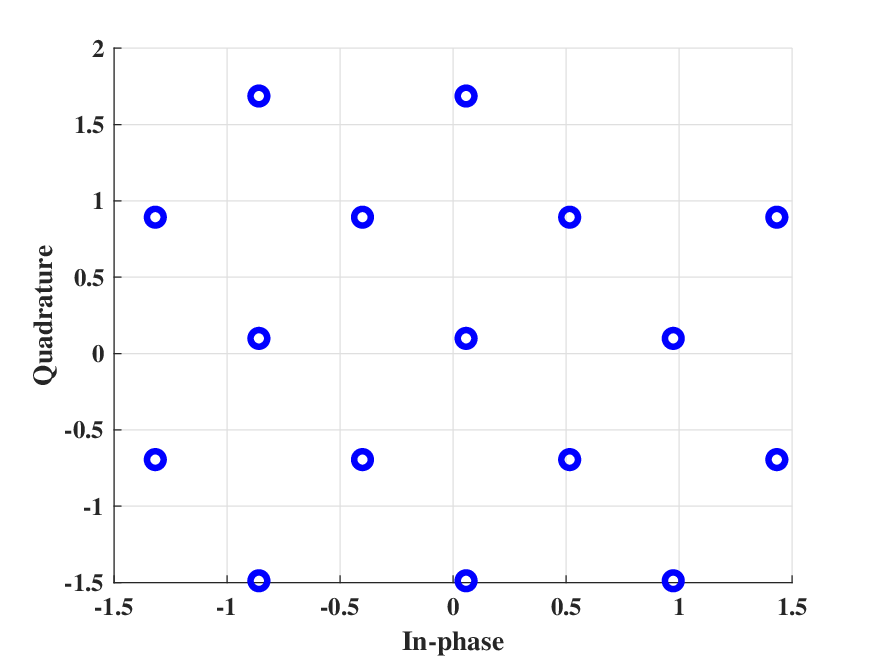}
  \caption*{Optimum, M=16}
\end{minipage}
\begin{minipage}[b]{0.22\textwidth}
  \centering
  \includegraphics[width=3.2cm,height=2.4cm]{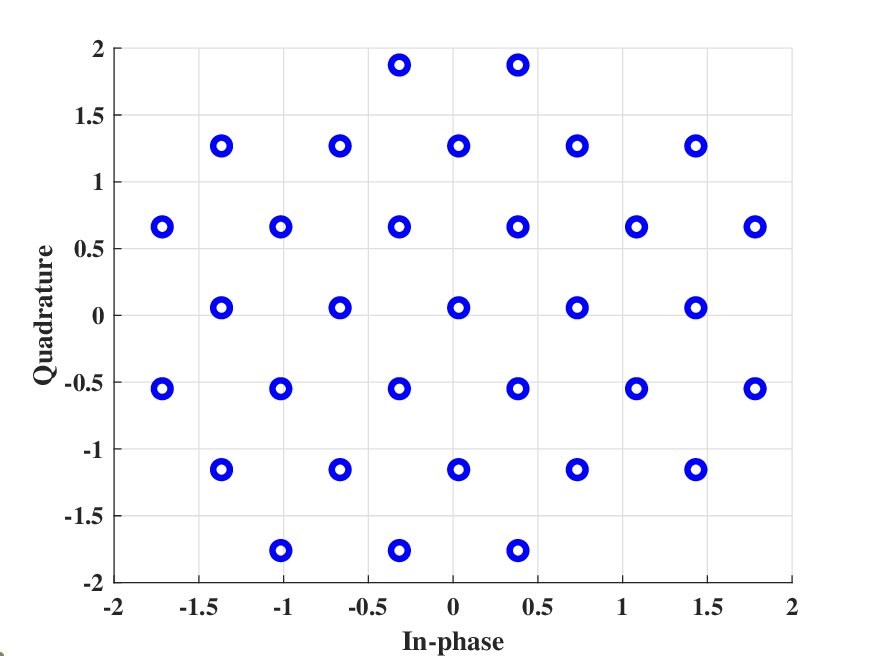}
  \caption*{Optimum, M=32}
\end{minipage}
\begin{minipage}[b]{0.22\textwidth}
  \centering
  \includegraphics[width=3.2cm,height=2.4cm]{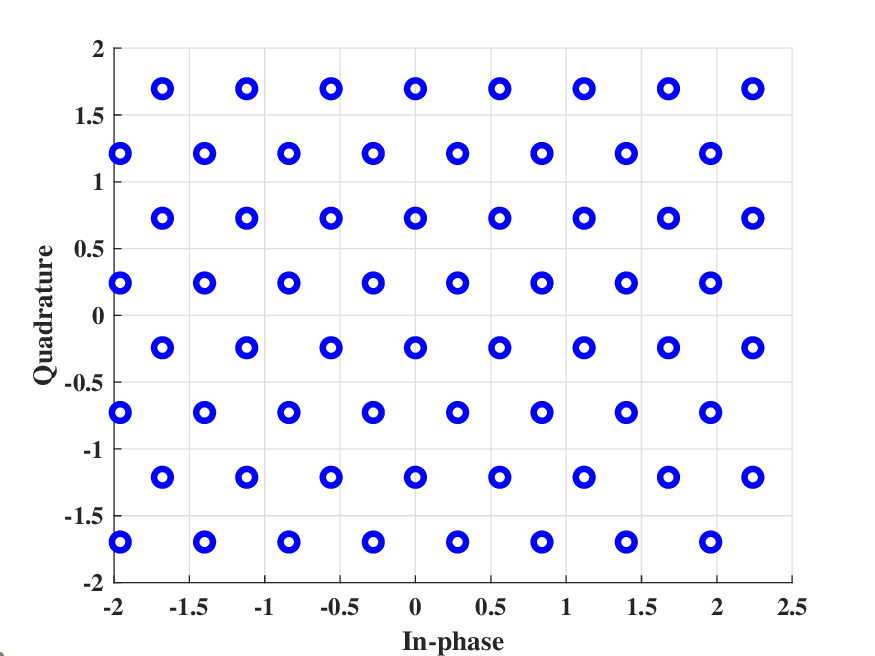}
  \caption*{Optimum, M=64}
\end{minipage}
\vspace{1em} 

\begin{minipage}[b]{0.22\textwidth}
  \centering
  \includegraphics[width=3.4cm,height=2.5cm]{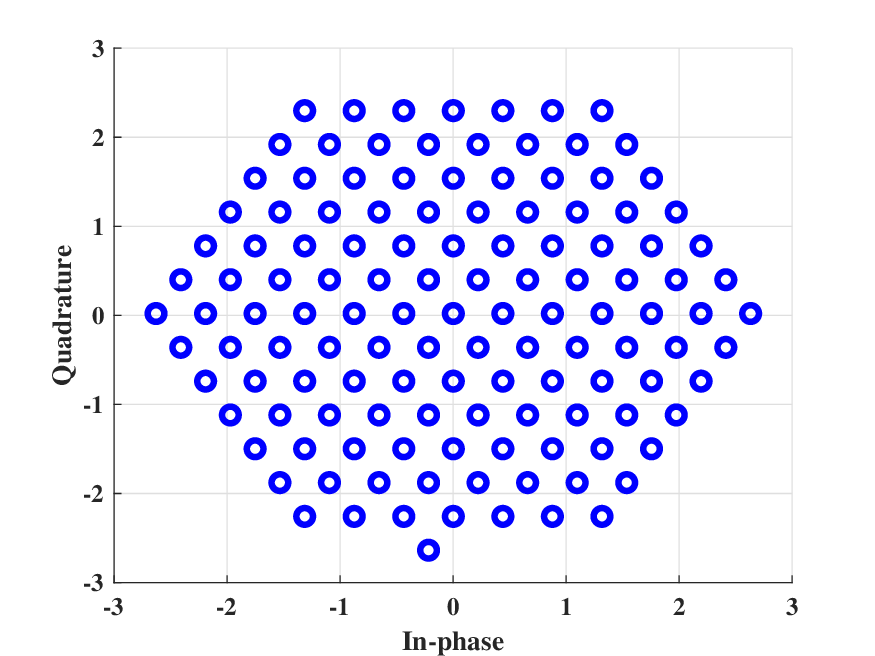}
  \caption*{Optimum, M=128}
\end{minipage}
\begin{minipage}[b]{0.22\textwidth}
  \centering
  \includegraphics[width=3.4cm,height=2.5cm]{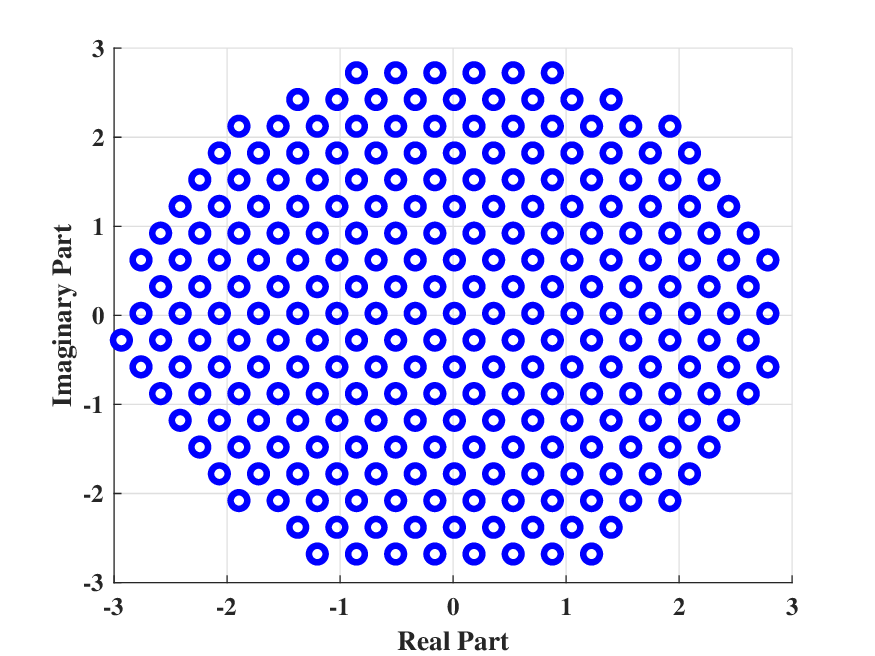}
  \caption*{Optimum, M=256}
\end{minipage}
\begin{minipage}[b]{0.22\textwidth}
  \centering
  \includegraphics[width=3cm,height=2.4cm]{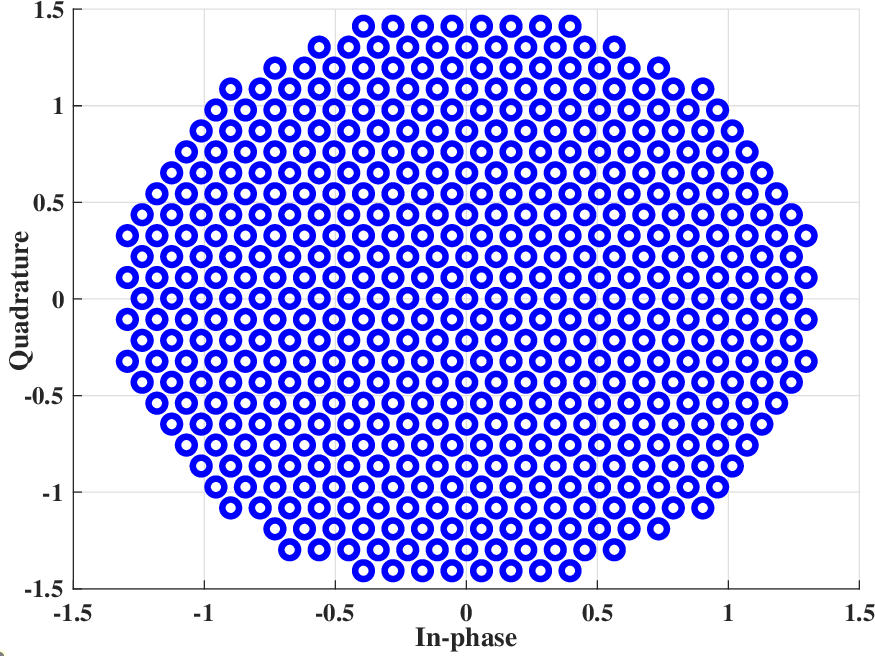}
  \caption*{Optimum, M=512}
\end{minipage}
\begin{minipage}[b]{0.22\textwidth}
  \centering
  \includegraphics[width=3cm,height=2.4cm]{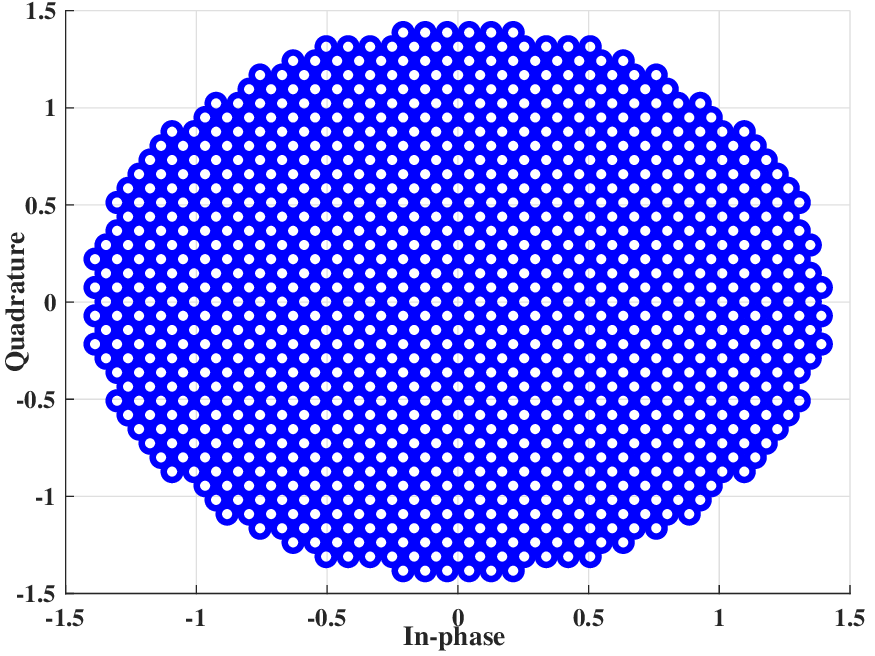}
  \caption*{Optimum, M=1024}
\end{minipage}
\vspace{1em}
\caption{M-ary constellation from DR of 3D-to-2D.}
\label{fig:2}
\end{figure}
\begin{figure}[!t]
\centering
\includegraphics[width=5.3cm,height=4cm]{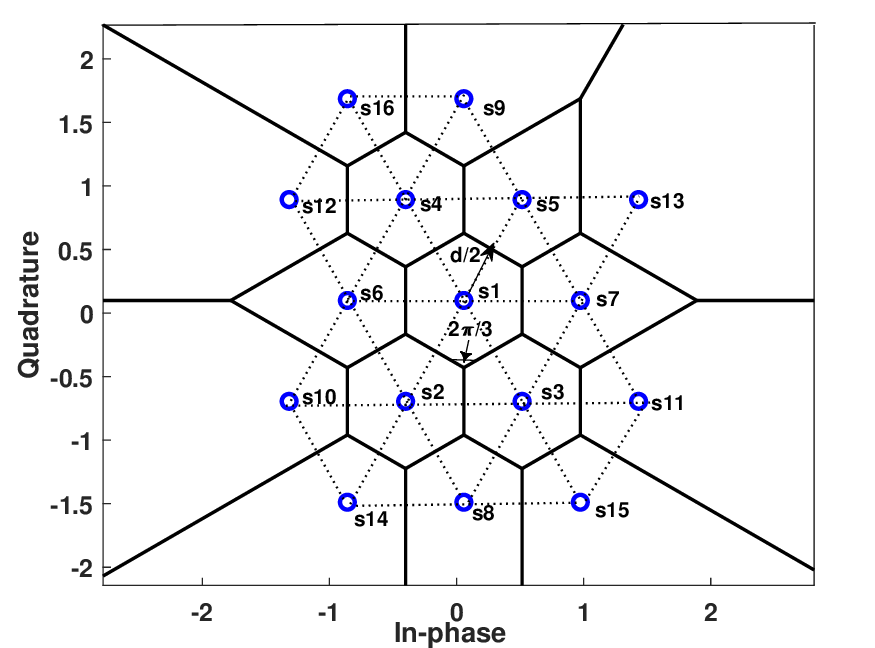}
\caption{16-ary constellation from DR of 3D-to-2D.}
\label{fig:3}
\end{figure}
\section{Design of Advanced 3D Constellations} 
In 3D signal space, a constellation of signal points is defined as a set with a uniform distribution. 
In this work, we design a 3D-HQAM constellation that extends the classical 2D-HQAM structure to $\mathbb{R}^3$ to achieve improved spatial separation and error performance. The constellation is defined as a finite set $\mathcal{S}_M = \{\mathbf{s}_i \in \mathbb{R}^3 : i = 1,2, \dots, M\}$ containing $M$ symbols,  where $\mathcal{S}_M$ is a subset of the infinite grid $\ {\mathcal{G}} = \{{v\in{\mathbb{R}^3}}:v =c_1v_1+c_2v_2  +c_3v_3+z_0\}$ with $v_1 = [\sqrt{2}d/3,\sqrt{6}d/3, d/3]^T$, $v_2 = [2\sqrt{2}d/3 ,0, d/3]^T$, and $v_3 = [-\sqrt{2}d/3, -\sqrt{6}d/3, d/3]^T$ as the basis vectors of the 3D space \cite{carle2000topological}. Here, \(d > 0\) denotes the smallest distance between points. Where \(c_1\), \(c_2\), \(c_3\) are integer coefficients $\mathbb{I}$ and \(z_0\) $\in \mathbb{R}^3$ is an offset of the grid. The basis vectors form a dense and approximately uniform spatial network, allowing efficient symbol packing while preserving a maximum MED between neighboring points. The constellation points \(\mathcal{S}_M\) are illustrated in Fig.~\ref{fig:1}. The key benefit of 3D constellation points is their ability to mitigate the reduction in MED in higher modulation orders, which otherwise degrades noise resistance. 

In 3D modulation, constellation points are distributed in a 3D space, which increases the Euclidean distance between them relative to one-dimensional (1D) and 2D schemes. This generally results in a reduced SEP. When the MED is identical across modulation schemes, their SEP performance is comparable. However, the 3D space allows for denser packing of constellation points, thereby improving spectral efficiency. Table~\ref{tab_1} compares the MEDs of the proposed 3D-HQAM constellations with their 2D-HQAM \cite{Sukhsagar2025accurate}, with all constellations normalized to the average unit power. Under the constraint $\text{MED} = 1$, a constellation with a lower total energy $E_s$ indicates better performance. This normalization facilitates analysis of the 3D point distributions using geometric methods. For a given modulation order $M$, a lower $E_s$ implies a higher MED, thus enhancing overall performance~\cite{chen2015design}.
As summarized in Table~\ref{tab_1}, the MED of 3D-HQAM increases with modulation order. For example, the 3D constellation achieves a 66.65\% higher MED than its 2D counterpart for 64-HQAM \cite{Sukhsagar2025accurate}. This improvement increases to 160.81\% for 1024-HQAM, indicating significant gains in error performance for higher-order 3D constellations.
\begin{table}[htbp]
\begin{center}
\scriptsize 
\caption{The MEDs of 2D \cite{Sukhsagar2025accurate} and 3D, HQAM Constellation }
\label{tab_1}
\begin{tabular}{|c|c|c|c|}
\hline
$M$ & 2D \cite{Sukhsagar2025accurate} & 3D & Increase (\%) \\
\hline
8 & 0.9428 & 1.0573 & 12.14 \\
\hline
16 & 0.6666 & 0.91654 & 37.49 \\
\hline
32 & 0.47634 & 0.69943 & 46.83 \\
\hline
64 & 0.33588 & 0.55974 & 66.65 \\
\hline
128 & 0.23776 & 0.4384 & 84.39 \\
\hline
256 & 0.16837 & 0.34664 & 105.88 \\
\hline
512 & 0.11900 & 0.27552 & 131.53 \\
\hline
1024 & 0.084047 & 0.2192 & 160.81 \\
\hline
\end{tabular}
\end{center}
\end{table}
\section{SEP Approximation for AWGN Channels}
Let $\mathbf{x} = (x_1, x_2, x_3)$ represent a transmitted signal vector chosen from a 3D constellation $\mathcal{S}_M \subset \mathbb{R}^3$. The transmission is assumed to occur over an AWGN channel. Consequently, the received vector $\mathbf{y} = (y_1, y_2, y_3)$ is given by
\begin{equation}
\label{eqn_1}
\mathbf{y} = \mathbf{x} + \mathbf{w},
\end{equation}
where $\mathbf{w} = (w_1, w_2, w_3)$ is a noise vector whose components are independent and identically distributed Gaussian random variables with zero mean and two-sided power spectral density $N_0/2$.
Given the transmission of symbol $\mathbf{x}$, the received vector $\mathbf{y}$ is modeled as a Gaussian random vector centered at $\mathbf{x}$. The conditional joint probability density function of $\mathbf{y}$ given $\mathbf{x}$ is defined as \cite{khabbazian2009exact}
\begin{equation}
\label{eqn_2}
p_{\mathbf{y}}(\mathbf{y} | \mathbf{x}) = \frac{1}{(2\pi\sigma^2)^{3/2}} \exp\left( -\frac{d^2}{2\sigma^2} \right),
\end{equation}
where $d$ denotes the Euclidean distance between $\mathbf{y}$ and $\mathbf{x}$, and \(\sigma^2 = N_0/2\) is noise variances of the transmitted signal.
To address the increased complexity of decision of 3D constellations, which arises due to a larger number of constellation points compared to 2D, a DR \cite{fodor2002survey} approach is used by exploiting the mapping relationship between the 3D and 2D coordinates \cite{zhao2024262144}. 

In this paper, the 3D-HQAM constellation \( \mathbf{X} \in \mathbb{R}^{M \times 3} \) is projected onto a 2D plane using Particle Swarm Optimization (PSO) \cite{kramer2013dimensionality} while preserving the MED. The objective is to find an optimal projection matrix \( \mathbf{P} \in \mathbb{R}^{3 \times 2} \), which minimizes the deviation between the projected MED and the target MED \( \Delta_{\text{target}} \), formulated as:
\[
\min_{\mathbf{P} \in \mathbb{R}^{3 \times 2}} \left( d_{\min}(\mathbf{X} \cdot \mathbf{P}) - \Delta_{\text{target}} \right)^2,
\]
where \( d_{\min}(\cdot) \) denotes the MED among the projected constellation points.
Each particle in PSO corresponds to a candidate projection matrix \( \mathbf{P} \in \mathbb{R}^{3 \times 2} \). During optimization, the particle positions are iteratively updated based on both personal best and global best solutions, until convergence is achieved.
 The search process is governed by the inertia weight \(w\) and the acceleration coefficients \( d_1, d_2 \in [l_b, u_b] \). After convergence, the optimal projection matrix \( \mathbf{P} \) maps each 3D point \( \mathbf{x}_j \) to a 2D complex point \( s_j = x_j + i y_j \). The reduced 2D constellation preserves the distance properties of the 3D constellation and is further used for SEP evaluation, as depicted in Fig.~\ref{fig:2} and in Algorithm~\ref{alg:1}.
\begin{algorithm}
\caption{Dimensionality Reduction of 3D-HQAM via PSO}
\label{alg:1}
\begin{algorithmic}[1]
\REQUIRE 3D constellation $\mathbf{X} \in \mathbb{R}^{M \times 3}$, target MED $\Delta_{\text{target}}$, number of particles $n$, number of iterations $T$, PSO parameters \( (w, d_1, d_2, l_b, u_b) \)

\ENSURE 2D constellation $\{s_j\}_{j=1}^{M}$ where $s_j = x_j + i y_j$
\STATE Initialize $n$ particles $\mathbf{p}_i \in \mathbb{R}^{3 \times 2}$ uniformly in $[l_b, u_b]$
\STATE Set velocity $\mathbf{v}_i = \mathbf{0}$, personal best $\mathbf{p}_i^{\text{best}} = \mathbf{p}_i$
\STATE Evaluate fitness $f_i$ using the Objective function (distance error)
\STATE Set global best $\mathbf{g}^{\text{best}} = \arg\min f_i$
\FOR{$t = 1$ to $T$}
    \FOR{$i = 1$ to $n$}
        \STATE Update $\mathbf{v}_i$ and $\mathbf{p}_i$ using PSO update rules
        \STATE Clip $\mathbf{p}_i$ within $[l_b, u_b]$
        \STATE Evaluate $f_i$, update $\mathbf{p}_i^{\text{best}}$ and $\mathbf{g}^{\text{best}}$ if improved
    \ENDFOR
\ENDFOR
\STATE Reshape $\mathbf{g}^{\text{best}} \rightarrow \mathbf{P} \in \mathbb{R}^{3 \times 2}$
\STATE Project: $\mathbf{Y} = \mathbf{X} \cdot \mathbf{P}$, form $s_j = x_j + i y_j$
\RETURN Reduced constellation $\{s_j\}$
\end{algorithmic}
\end{algorithm}

The maximum likelihood (ML) decision regions, depicted in Fig.~\ref{fig:3}, central areas that form equilateral hexagons with internal angles of \(2\pi/3\). The outer regions may include one or more angles of \(2\pi/3\). 
An approximation \cite{Sukhsagar2025accurate} of the SEP for an \(M\)-ary signal in an AWGN channel is given by 

\begin{equation}
\label{eqn_3}
  P_{NN} = K Q\left(\frac{d}{2\sigma}\right),
\end{equation}
where \( Q(\cdot) \) represents the Gaussian \( Q \)-function, defined as \( Q(z) = \frac{1}{\sqrt{2\pi}} \int_{z}^{\infty} e^{-\frac{t^2}{2}} \, dt \). Here, \( K \) represents the average number of NNs, calculated as \( K = \frac{1}{M} \sum_{i=0}^{M-1} K(i) \), where \( K(i) \) is the count of NNs for the symbol \( s_i \).

Fig.~\ref{fig:3} shows six distinct boundary regions for 16-HQAM constellations. The region of $s1$ includes six inner neighboring symbols (s2, s3, s4, s5, s6, s7). The boundary of $s5$ covers five corner symbols (s1, s4, s7, s9, s13). The region of $s8$ includes four symbols (s2, s3, s14, s15) at the top and bottom. The boundary of $s{13}$ includes two symbols (s5, s7) on the left and right. Finally, the regions of $s{11}$ and $s{14}$ include three symbols (s3, s7, s15) and (s2, s8, s10) on the left, right, and top sides, respectively. The correct decision probability is the product of probabilities keeping the symbol within its boundary, while the error probability is the product of noise probabilities causing it to fall outside the boundary \cite{singha2021error, rugini2016symbol}.
As seen in Fig.~\ref{fig:3}, there are 4 regions of type $s1$, 3 regions of type $s5$, 1 region of type $s8$, 1 region of type $s{13}$, 4 regions of type $s{11}$, and 3 regions of type $s{14}$.
By symmetry, the weighted sum of correct decision probabilities for each region type is given as
\begin{equation}
\label{eqn_4}
P_c^{16-HQAM} = 4P_{c,s1} + 3P_{c,s5} + P_{c,s8} + P_{c,s13} + 4P_{c,s11} + 3P_{c,s14}
\end{equation}
The correct decision probability for each region depends on the number of independent noise components \cite{singha2021error}, which can be expressed as follows:
\begin{equation}
\label{eqn_5}
P_{c,s1} = \left( 1 -Q\left( \frac{d}{2\sigma} \right) \right)^6
\end{equation}
\begin{equation}
\label{eqn_6}
P_{c,s5} = \left( 1 - Q\left( \frac{d}{2\sigma} \right) \right)^5
\end{equation}

\begin{equation}
\label{eqn_7}
P_{c,s8} = \left( 1 - Q\left( \frac{d}{2\sigma} \right) \right)^4
\end{equation}

\begin{equation}
\label{eqn_8}
P_{c,s13} = \left( 1 - Q\left( \frac{d}{2\sigma} \right) \right)^2
\end{equation}

\begin{equation}
\label{eqn_9}
P_{c,s11} = \left( 1 - Q\left( \frac{d}{2\sigma} \right) \right)^3
\end{equation}

\begin{equation}
\label{eqn_10}
P_{c,s14} = \left( 1 - Q\left( \frac{d}{2\sigma} \right) \right)^3
\end{equation}

Substituting (\ref{eqn_5})--(\ref{eqn_10})
 into (\ref{eqn_4}), the average correct decision probability can be expressed as
\begin{align}\label{eqn_11}
&P_c^{16-HQAM} = \frac{1}{16} \Bigg[ 
 4 \left( 1 - Q\left( \frac{d}{2\sigma} \right) \right)^6 + 3 \left( 1 - Q\left( \frac{d}{2\sigma} \right) \right)^5 \notag \\ 
& + \left( 1 - Q\left( \frac{d}{2\sigma} \right) \right)^4 
+ 7 \left( 1 - Q\left( \frac{d}{2\sigma} \right) \right)^3 + \left( 1 - Q\left( \frac{d}{2\sigma} \right) \right)^2\Bigg],
\end{align}
Using the binomial expansion
\((1 - a)^n = \sum_{k=0}^{n} (-1)^k \binom{n}{k} a^k\)
and applying the relationship \( P_e = 1 - P_c \). The SEP of 16-HQAM is given as
\begin{align}
\label{eqn_12}
P^{16-HQAM}_{e} =\ & \frac{33}{8} Q\left( \frac{d}{2\sigma} \right) 
- \frac{59}{8} Q^2\left( \frac{d}{2\sigma} \right) 
+ \frac{121}{16} Q^3\left( \frac{d}{2\sigma} \right) \notag \\ 
& - \frac{19}{4} Q^4\left( \frac{d}{2\sigma} \right) 
+ \frac{27}{16} Q^5\left( \frac{d}{2\sigma} \right) 
- \frac{1}{4} Q^6\left( \frac{d}{2\sigma} \right),
\end{align}
where \( \sigma = \sqrt{\frac{N_0}{2}} \), \( d = \sqrt{\frac{E_s}{ 1.12}} \), and  \( \gamma = \frac{E_s}{N_0} \).
Using a similar approach as in (\ref{eqn_12}), the SEP for other HQAM constellations can be obtained. Hence, the generalized SEP expression for an arbitrary $M$-ary HQAM scheme can be proposed as

\begin{align}
\label{eqn_13}
P_e^{M-HQAM} = \sum_{n=1}^{6} (-1)^{n+1} b_n Q^n\left( \frac{\gamma}{A} \right).
\end{align}
The values of $A$ and $b_n$ depend on the constellation order $M$ and are provided in Table~\ref{tab_2}. By substituting these values into (\ref{eqn_13}), the SEP expression for various HQAM constellation orders can be obtained.
\begin{table}[htbp]
\begin{center}
\scriptsize 
\centering
\caption{Parameters for Different Modulation Orders}
\label{tab_2}
\begin{tabular}{|c|c|c|c|c|c|c|c|}
\hline
$M$ & $A$ & \multicolumn{6}{c|}{$b_n$ values} \\
\cline{3-8}
     &     & $b_1$ & $b_2$ & $b_3$ & $b_4$ & $b_5$ & $b_6$ \\
\hline
16   & 2.24   & $\frac{33}{8}$  & $\frac{59}{8}$   & $\frac{121}{16}$ & $\frac{19}{4}$  & $\frac{27}{16}$ & $\frac{1}{4}$ \\
\hline
32   & 4.02   & $\frac{75}{16}$ & $\frac{305}{32}$ & $\frac{349}{32}$ & $\frac{237}{32}$& $\frac{89}{32}$ & $\frac{7}{16}$ \\
\hline
64   & 6.32   & $\frac{161}{32}$ & $\frac{349}{32}$   & $\frac{209}{16}$ & $\frac{291}{32}$  & $\frac{93}{32}$ & $\frac{9}{16}$ \\
\hline
128  & 10.18  & $\frac{43}{8}$  & $\frac{393}{32}$   & $\frac{979}{64}$ & $\frac{1403}{128}$  & $\frac{137}{32}$ & $\frac{91}{128}$ \\
\hline
256  & 16.26  & $\frac{711}{128}$  & $\frac{3345}{256}$   & $\frac{4265}{256}$ & $\frac{3105}{256}$  & $\frac{1221}{256}$ & $\frac{101}{128}$ \\
\hline
512  & 26.54  & $\frac{1455}{256}$  & $\frac{6977}{512}$   & $\frac{9025}{512}$ & $\frac{6633}{512}$  & $\frac{2621}{512}$ & $\frac{217}{256}$ \\
\hline
1024 & 41.14  & $\frac{5913}{1024}$  & $\frac{14361}{1024}$   & $\frac{18757}{1024}$ & $\frac{13875}{1024}$  & $\frac{2751}{512}$ & $\frac{57}{64}$ \\
\hline
\end{tabular}
\end{center}
\end{table}
\begin{figure}[!t]
\centering
\includegraphics[width=7.3cm,height=5.4cm]{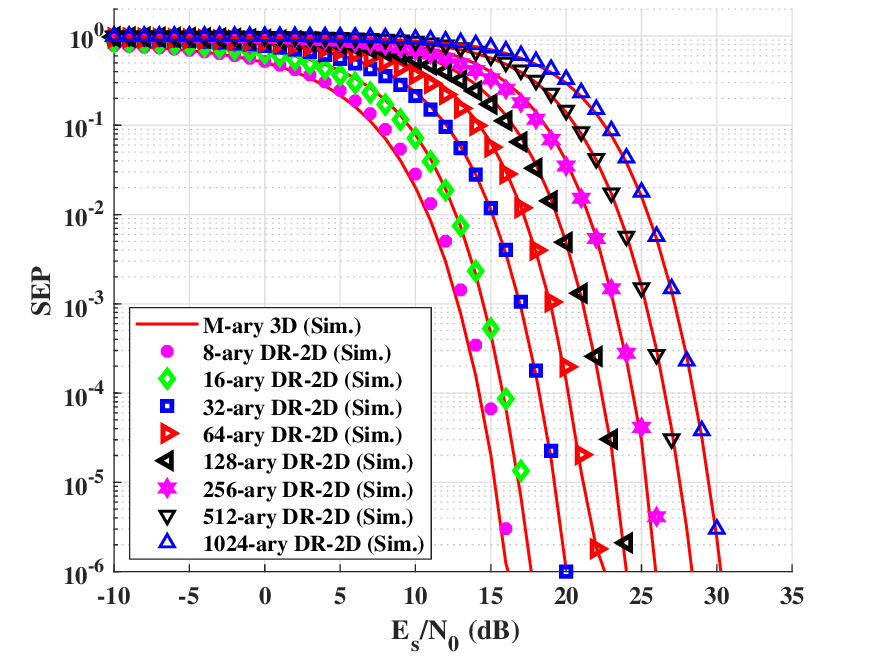}
\caption{Performance analysis over SEP vs. $E_s/N_0$ for 3D HQAM and its after DR 2D-HQAM.}
\label{fig:4}
\end{figure}
\begin{figure}[!t]
\centering
\includegraphics[width=6.6cm,height=5.4cm]{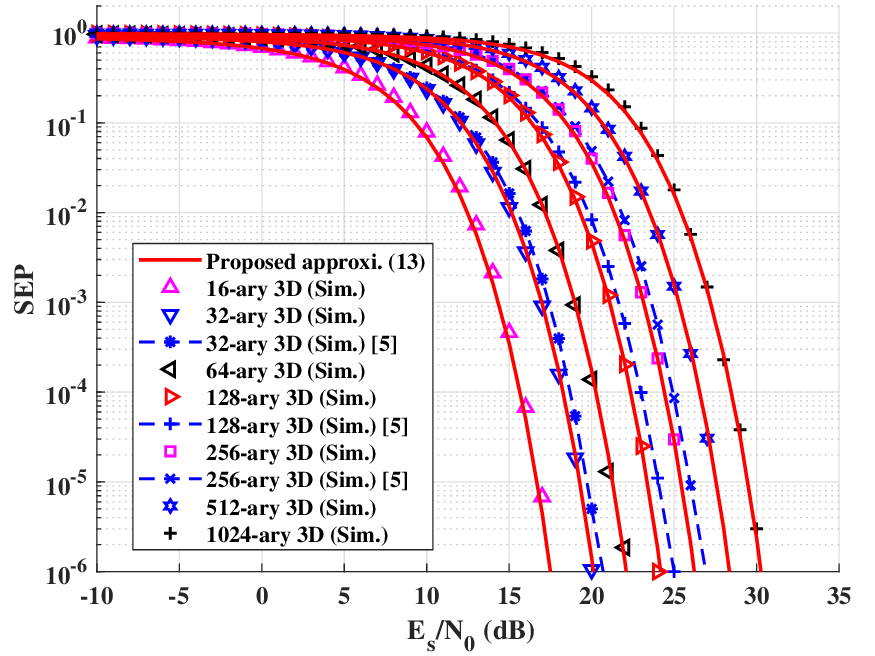}
\caption{Simulated and analytical SEP for $M$-HQAM in AWGN channel}
\label{fig:5}
\end{figure}
\begin{figure}[!t]
\centering
\includegraphics[width=7cm,height=5.4cm]{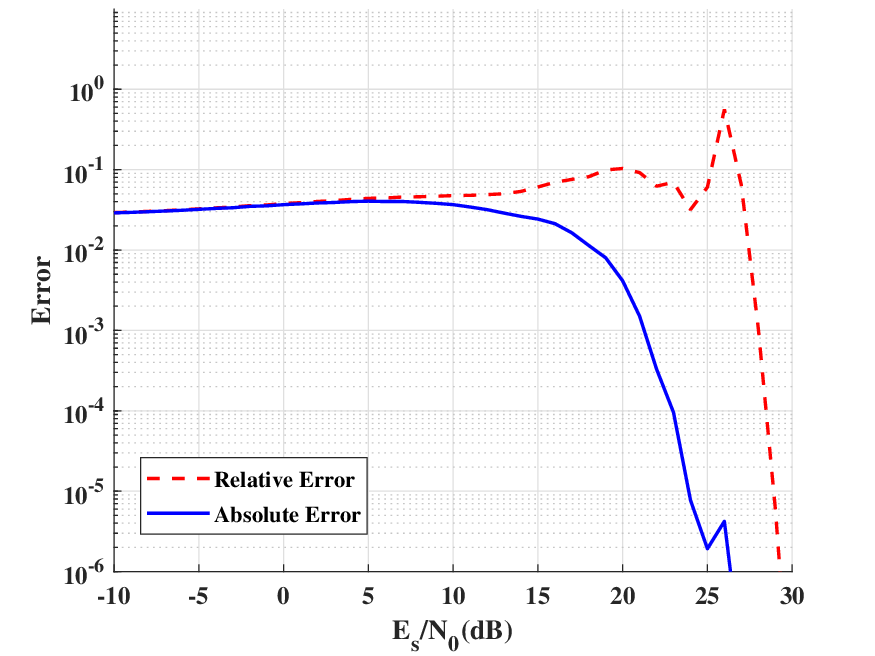}
\caption{Error vs received SNR of 256-ary 3D-HQAM}
\label{fig:6}
\end{figure}


\section{Performance Analysis}
Fig.~\ref{fig:4} presents the SEP comparison between the 3D-HQAM and the DR-based 2D-HQAM constellation points. The SEP curves indicate that the reduced 2D-HQAM closely follows the performance of the 3D-HQAM across all constellation orders. This demonstrates that the reduction method effectively preserves the error performance for all constellation sizes $M$, making it a suitable and reliable approach for analyzing the performance of 3D-HQAM systems.

Fig.~\ref{fig:5} illustrates the SEP performance obtained from eq. (\ref{eqn_13}) against $E_{s}/N_{0}$ for 3D-HQAM constellations over an AWGN channel. The results show a strong agreement between the theoretical and simulated SEP curves across all SNR levels and constellation sizes, validating the accuracy of the proposed approximation. Furthermore, the performance is compared with previously reported approximations in \cite{kang2011construction}, highlighting the effectiveness of the proposed expression. We set \( P_e = 10^{-6} \) as the reference performance level for comparison. Across all SNRs, the proposed 3D-ary system outperforms the design presented in \cite{kang2011construction}. For the $32$-ary, $128$-ary, and $256$-ary constellations, the proposed method demonstrates improvements of approximately $0.70$ dB, $1.0$ dB, and $1.1$ dB, respectively, over the results in \cite{kang2011construction}. These findings highlight the enhanced efficiency and precision of the proposed approach.

Fig.~\ref{fig:6} illustrates the absolute error (AE) and relative error (RE) for 3D 256-HQAM over an AWGN channel, defined as
\(
\text{AE} = |P - P_{\text{exact}}|,
\)
\(
\text{RE} = {|P - P_{\text{exact}}|}/{P_{\text{exact}}}\),
where \( P \) denotes the SEP obtained from proposed expression (\ref{eqn_13}) and \( P_{\text{exact}} \) represents the exact SEP obtained from simulation. The AE consistently decreases with increasing \( E_s/N_0 \), falling below \(10^{-5}\) beyond 22~dB, indicating strong agreement between simulation and theory. The RE represents uncertainty in the approximation with respect to the exact SEP value. The proposed approximation shows high precision, with the RE remaining below \( 10^{-1} \) (less than 1\%) in the SNR range of -$10$ dB to $25$ dB, with a slight increase above $25$~dB due to the small magnitude of the SEP values. These results confirm the model’s accuracy, particularly for \( E_s/N_0 \geq 10 \)~dB.

\begin{figure*}[!t] 
    \centering
    \begin{subfigure}[b]{0.45\textwidth}
        \centering
        \includegraphics[width=6.5cm,height=5.2cm]{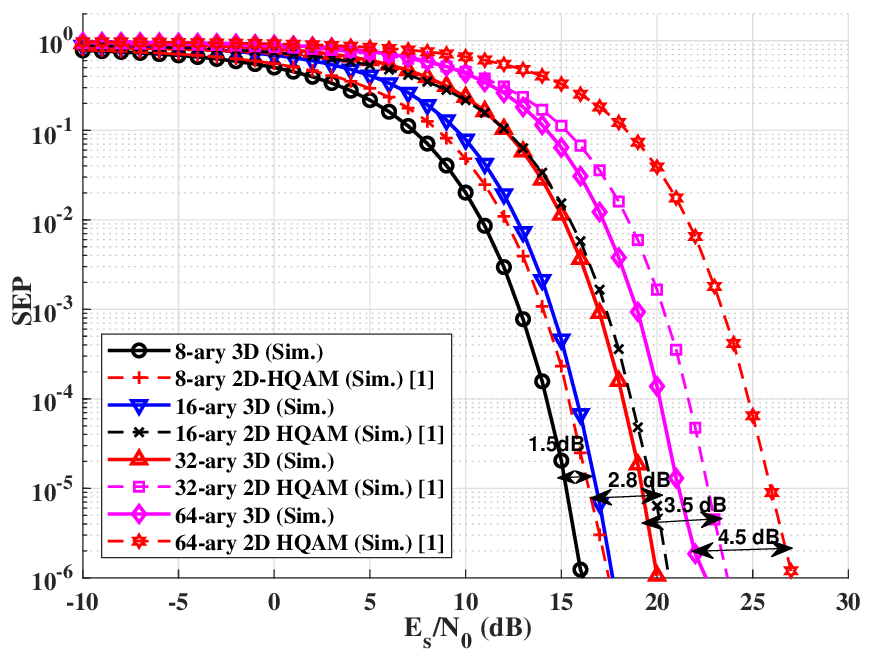}
        \caption{}
        \label{fig:7a}
    \end{subfigure}
    \hfill
    \begin{subfigure}[b]{0.45\textwidth}
        \centering
        \includegraphics[width=6.5cm,height=5.2cm]{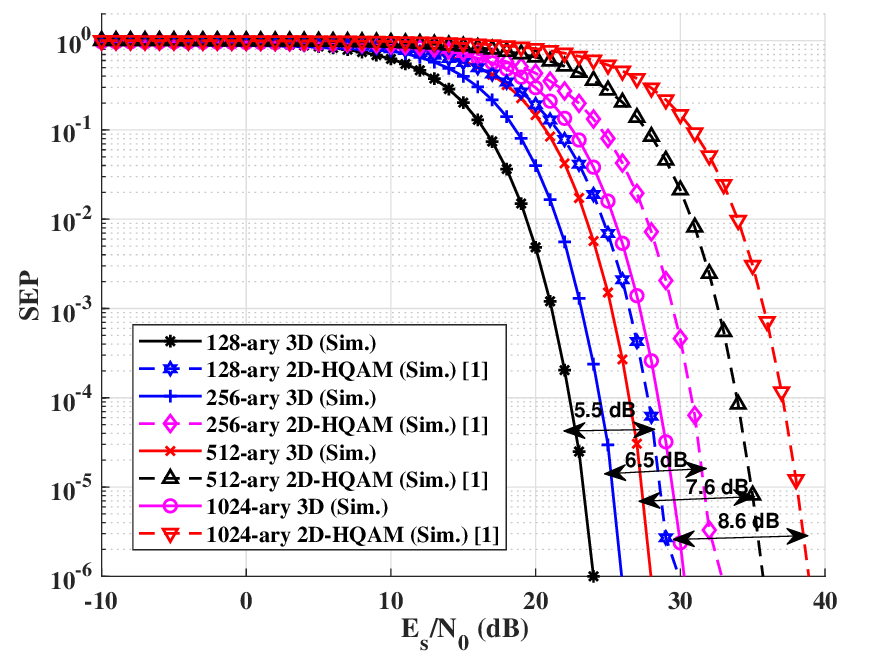}
        \caption{}
        \label{fig:7b}
    \end{subfigure}
    
    \caption{Performance comparison of 3D-HQAM and 2D-HQAM schemes.}
    \label{fig:7}
\end{figure*}

To assess the effect of MED on error performance, Fig.\ref{fig:7} compares the SEP of the proposed 3D-HQAM constellations with the 2D-HQAM scheme presented in \cite{Sukhsagar2025accurate}. Using SNR gains at an SEP level of $10^{-5}$, as a reference, the 3D-HQAM consistently demonstrates superior performance over 2D-HQAM across all SNR values. Specifically, the results indicate notable SNR gains of 1.5 dB, 2.8 dB, 3.5 dB, and 4.5 dB for constellation sizes of 8-ary, 16-ary, 32-ary, and 64-ary, respectively. When the constellation order is increased to 128-ary, 256-ary, 512-ary, and 1024-ary, the 3D-HQAM schemes offer performance gains of approximately 5.5 dB, 6.5 dB, 7.6 dB, and 8.6 dB, respectively, compared to their 2D counterparts. These findings confirm superiority of the proposed 3D-HQAM over the 2D-HQAM scheme in \cite{Sukhsagar2025accurate, rugini2016symbol} and their consistency with the MED analysis in Table~\ref{tab_1}.

\section{Conclusion}
This paper presents an efficient method for constructing higher-order 3D signal constellations for digital communication systems. The proposed approach expands the conventional 2D-HQAM into a 3D-HQAM signal space, resulting in structured 3D lattice constellations. To handle the increased decision complexity caused by the larger number of constellation points, a DR method is applied. This enables the derivation of closed-form SEP expressions under AWGN conditions. Simulation results validate the accuracy of the theoretical SEPs. The proposed 3D constellations also exhibit significantly larger MEDs compared to 2D-HQAM. As a result, they achieve much lower SEPs, improving communication reliability. These advantages make 3D-HQAM a strong candidate for next-generation high-quality communication systems. The proposed model should also be benchmarked against recent advances in multidimensional signal transmission.

\bibliography{CL}

\bibliographystyle{IEEEtran}

\vfill

\end{document}